      [1999/12/01 v1.4c Il Nuovo Cimento]
\documentclass{cimento}

\usepackage{epsfig}
\usepackage{bm,amsmath,amssymb}

\def\beq{\begin{equation}}
\def\eeq{\end{equation}}
\def\beqr{\begin{eqnarray}}
\def\eeqr{\end{eqnarray}}
\def\bdpm{\begin{displaymath}}
\def\edpm{\end{displaymath}}

\title{Effective field theory for top physics}
\author{ 
P.~Ko \from{ins:y}
}
\instlist{%\inst{ins:x} Xavier's school for gifted youngsters - Westchester,N.Y.
  \inst{ins:y} School of Physics, KIAS, Seoul 130-722, Korea,
  }
\PACSes{\PACSit{14.65.Ha}{Top quarks.}
\PACSit{12.60.-i}{Models beyond the standard model}}
\begin{document}

\maketitle

\begin{abstract} 
We study the top forward-backward asymmetry (FBA $\equiv A_{\rm FB}$)  
reported by CDF and D0 Collaborations in the effective lagrangian approach. 
Using dimension-6 effective largrangians for  
$q \bar{q} \rightarrow t \bar{t}$, we study the $t\bar{t}$ production 
cross section and the $A_{\rm FB}$, and a few observables:     
the FB spin-spin correlation that is strongly 
correlated with the $A_{\rm FB}$, and longitudinal top polarization
as a new probe of chiral structures for possible new physics scenarios.
\end{abstract}

\section{Introduction}

Top quark is the heaviest particle observed so far, and could be 
sensitive to the underlying physics of electroweak symmetry breaking. 
So far there is no clear deviation from the SM predictions, 
possibly except for the $A_{\rm FB}$ observed at the Tevatron:
\begin{equation}
A_{\rm FB} ({\rm CDF}) = ( 0.158 \pm 0.074 )  , \ \
A_{\rm FB} ({\rm D0}) =   ( 0.196 \pm 0.065 ) ,
\end{equation}
compared with the SM prediction $A_{\rm FB}\sim 0.078$ 
\cite{Antunano:2007da}. 
This $\sim 1-2\sigma$ deviation and the mass-dependent $A_{\rm FB}$ 
might be due to some new physics.
On the other hand, search for a new resonance decaying into  
$t\bar{t}$ pair has been  carried out  at the Tevatron. 
As of now, there is no clear signal for such a new resonance.  
And the $t\bar{t}$ production cross section is well described by the SM. 
Therefore, in this talk, I will consider the case where a new physics scale 
relevant to $A_{\rm FB}$ is large enough so that  production of a new 
particle is beyond the reach of the Tevatron \cite{ko1,ko2,ko3}. 
Then it is adequate to integrate out the heavy fields, and use the 
resulting effective lagrangian in order to study new physics effects 
on $\sigma_{t\bar{t}}$ and $A_{\rm FB}$ in a model independent way.  
At the Tevatron, the $t\bar{t}$ production is dominated by $q\bar{q} 
\rightarrow t\bar{t}$, and it would be sufficient to consider dimension-6
four-quark operators (the so-called contact interaction terms) 
to describe the new physics effects on the $t\bar{t}$ production 
at the Tevatron.  Note that a similar approach was adopted for the dijet 
production to constrain the composite scale of light quarks for long time.  
Before proceeding to the main topics of this talk, let us recall 
that similar approaches were taken in Refs.~
\cite{AguilarSaavedra:2010zi,Zhang:2010dr,Degrande:2010kt}. 
Also, there are effective field theory approaches for the same sign 
top pair production \cite{Degrande:2011rt}. 

\section{Effective field theory for top physics and new physical observables }

Our starting point is the effective lagrangian with dimension-6 
operators relevant to the $t\bar{t}$ production at the Tevatron \cite{ko1}: 
\begin{equation}
\mathcal{L}_6 = \frac{g_s^2}{\Lambda^2}\sum_{A,B}
\left[C^{AB}_{1q}(\bar{q}_A\gamma_\mu  q_A)(\bar{t}_B\gamma^\mu t_B)  + 
C^{AB}_{8q}(\bar{q}_A T^a\gamma_\mu q_A)(\bar{t}_B T^a\gamma^\mu  t_B)\right]
\end{equation}
where $T^a = \lambda^a /2$, $\{A,B\}=\{L,R\}$, and 
$L,R \equiv (1 \mp \gamma_5)/2$ 
with $q=(u,d)^T,(c,s)^T$.  
Using this effective lagrangian, we calculate the cross section up to 
$O(1/\Lambda^2)$, keeping only the interference term between 
the SM and new physics contributions.

We make one comment: the chromomagnetic operators of dim-5 would 
be generated at one loop level, whereas the $q\bar{q} \rightarrow t \bar{t}$
operators can be induced at tree level. 
Therefore the chromomagnetic operators will be suppressed further by 
$g_s / ( 4 \pi )^2 \times (loop \ function)$, compared with the dim-6 
operators we consider in this talk. Therefore we will ignore chromomagnetic
operators in this talk. 
(See Ref.~\cite{AguilarSaavedra:2010zi,Zhang:2010dr,Degrande:2010kt} 
for the discussion on this operator.)

Neglecting the transverse polarizations, the squared amplitude 
summed (averaged) over the final (initial) colors is given by 
%an expansion of the trace in Eq.~(\ref{eq:trace}) leads to 
\begin{equation}
\overline{|{\mathcal{M}}|^2}  =  \frac{g_s^4}{\hat{s}^2}\Bigg\{
{\cal D}_0  +
{\cal D}_1 (P_L+\bar{P}_L) 
%\nonumber \\ && \hspace{0.5cm} + 
+ {\cal D}_2 (P_L-\bar{P}_L)+
{\cal D}_3 P_L \bar{P}_L \Bigg\}\,.
\end{equation}
where $P_L$ and $\bar{P}_L$ are the longitudinal polarizations of 
$t$ and $\bar{t}$ \cite{ko2}. 

The unpolarized coefficient ${\cal D}_0$ leads to
the total cross section $\sigma_{t\bar{t}}$ and the forward-backward
asymmetry $A_{\rm FB}$.
On the other hand, the coefficient ${\cal D}_3$ gives the 
spin-spin correlations $C$ and $C_{\rm FB}$ considered 
and suggested before.
%\begin{widetext}
\begin{eqnarray}
{\cal D}_0  
& \simeq  & \frac{4}{9} \left\{
2 m_t^2 \hat{s} \left[
1+\frac{\hat{s}}{2\Lambda^2}\,(C_1+C_2)
\right] s_{\hat\theta}^2   \right. 
\\ 
& + & \left. 
\frac{\hat{s}^2}{2}\left[ \left(1+\frac{\hat{s}}{2\Lambda^2}\,(C_1+C_2)\right)
(1+c_{\hat\theta}^2)
+\hat\beta_t\left(\frac{\hat{s}}{\Lambda^2}\,(C_1-C_2)\right)c_{\hat\theta}
\right]\right\}   \nonumber 
\label{eq:ampsq}
\end{eqnarray}
%\end{widetext}
where $\hat{s} = (p_1 + p_2)^2$, $\hat\beta_t^2=1-4m_t^2/\hat{s}$,
and $s_{\hat\theta}\equiv \sin\hat\theta$ and 
$c_{\hat\theta}\equiv \cos\hat\theta$ with $\hat{\theta}$ being the polar
angle between the incoming quark and the outgoing top quark in the 
$t\bar{t}$ rest frame.  And the couplings are defined as:
$C_1 \equiv C_{8q}^{LL}+C_{8q}^{RR}$ and 
$C_2 \equiv C_{8q}^{LR}+C_{8q}^{RL}$. 
Since we have kept only up to the interference terms, there are 
no contributions from  the color-singlet operators with coupling 
$C_{1q}^{AB}$. 
The term linear in $\cos\hat{\theta}$ could
generate the forward-backward asymmetry 
which is proportional to $\Delta C \equiv (C_1 - C_2)$.
Note that both light quark and top quark should have chiral couplings
to the new physics in order to generate $A_{\rm FB}$ at the tree level
(namely $\Delta C \neq 0$).  This parity violation, if large, 
could be observed in the nonzero (anti)top spin polarization \cite{ko2}. 
The allowed region in the $(C_1,C_2)$ plane that is consistent with the 
Tevatron data  at the $1 \sigma$ level is 
%The allowed region is 
around $0.15 \lesssim C_1 \lesssim 0.97$ and $-0.67 \lesssim  C_2 
\lesssim -0.15$ for $\Lambda = 1$ TeV. 
The positive $C_1$ and the negative $C_2$ are preferred at the 
1 $\sigma$ level \cite{ko1,ko2}.   

Another interesting observable which is sensitive to the chiral 
structure of new physics affecting $q\bar{q} \rightarrow t\bar{t}$ 
is the top quark spin-spin correlation \cite{ko1}: 
\begin{equation}
C = \frac{\sigma(t_L\bar{t}_L + t_R\bar{t}_R) - 
\sigma(t_L\bar{t}_R + t_R\bar{t}_L)}{\sigma(t_L\bar{t}_L + t_R\bar{t}_R) + 
\sigma(t_L\bar{t}_R + t_R\bar{t}_L)} \,.
\end{equation}
Since new physics must have chiral couplings both to light quarks and 
top quark, the spin-spin correlation defined above will be affected.
In Ref.~\cite{ko1}, we proposed a new spin-spin FB asymmetry $C_{FB}$  
defined as $C_{FB} \equiv  C (\cos\theta \geq 0) -  C (\cos\theta \leq 0)$,
where $C(\cos\theta \geq 0 (\leq 0))$ implies the cross sections in the 
numerator of Eq.~(5) are obtained for the forward (backward) region: 
$\cos\theta \geq 0 (\leq 0)$.
In Ref.~\cite{ko1}, it was noticed that there is a clear strong correlation 
between $C_{FB}$ and $A_{FB}$. 
This correlation must be observed in the future measurements 
if the $A_{\rm FB}$ anomaly is real and  
a new particle is too heavy to be produced at the Tevatron.

\begin{figure}
\begin{center}
\includegraphics[width=6.0cm,height=6.0cm]{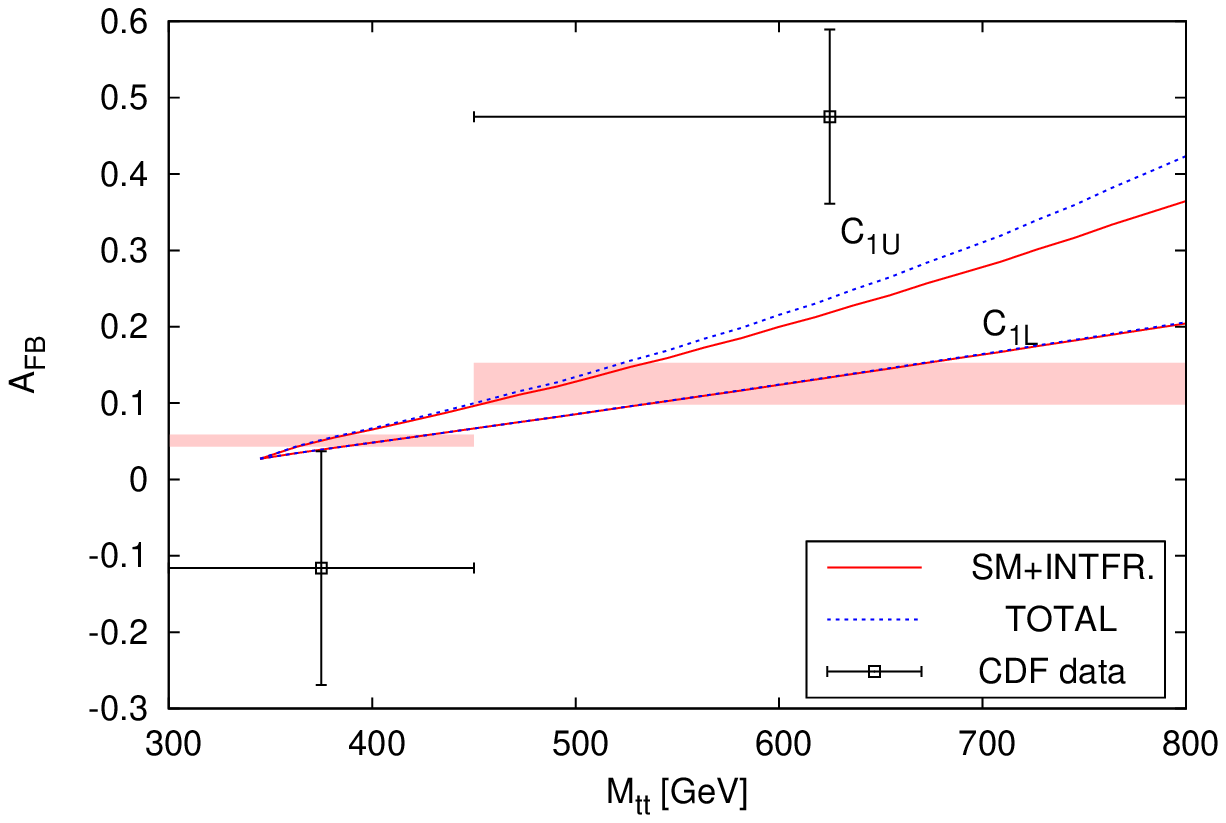} 
\includegraphics[width=6.0cm,height=6.0cm]{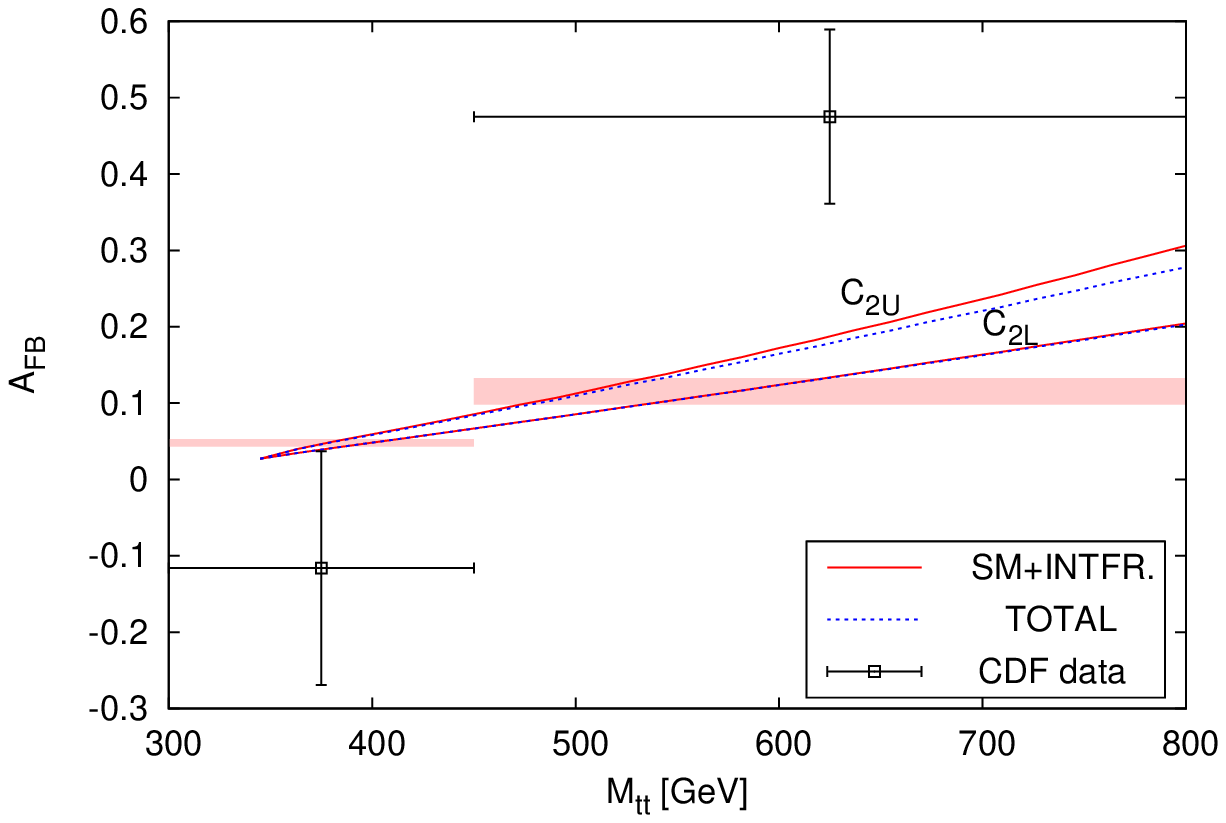}
\end{center}
\vspace{0.0cm}
\caption{\it 
Top FB asymmetry as functions of $M_{t\bar{t}}$. 
In the left frames we are taking
$C_1$ in the range between
$C_{1L}=0.15$ and $C_{1U}=0.97$ with $C_2=0$.
In the right frames, we vary $C_2$  in the range between
$C_{2L}=-0.15$ and $C_{2U}=-0.67$ with $C_1=0$. 
In each frame, the two bands are for  $A_{\rm FB}$ in the
lower and higher $M_{t\bar{t}}$ bins varying $C_1$ (left) and 
$C_2$ (right)  in the ranges delimited by $C_{1L,1U}$ and 
$C_{2L,2U}$, respectively,.
and the dots for the CDF data with errors.
In the solid (red) lines, we include only the SM contribution and
the one from the interference beween the SM and NP amplitudes while
the effects of $(NP)^2$ term have been added
in the dotted (blue) lines.
}
\label{fig:afb}
\end{figure}

One can also make predictions of the $A_{\rm FB}$  as functions of 
$M_{t\bar{t}}$ \cite{ko6}, which are shown in Fig.~2.  
Our results based on the effective lagrangian approach is significantly 
smaller than the CDF and D0 data. If this deviation is confirmed in the future 
analysis, it would imply that the effective lagrangian approach is not adequate 
to describe the top FB asymmetry at the Tevatron, and one has to consider 
various explicit models one by one, and investigate which
model describes all the data in a consistent way.

The other $P$-violating coefficient
${\cal D}_2$ could be observable at the Tevatron, revealing
genuine features of new physics responsible for $A_{\rm FB}$.
Explicitly, we have obtained 
\begin{equation}
{\cal D}_2 \simeq \frac{\hat{s}}{9\,\Lambda^2}\left[
(C_1^\prime + C_2^\prime) \hat{\beta_t} (1+c^2_{\hat{\theta}})
 +(C_1^\prime - C_2^\prime) (5-3 \hat{\beta}_t^2) c_{\hat{\theta}}\right]
\end{equation}
with $C_1^\prime  \equiv  C_{8q}^{RR}-C_{8q}^{LL}\,, \ 
C_2^\prime  \equiv  C_{8q}^{LR}-C_{8q}^{RL}.$
Therefore ${\cal D}_2$ will provide additional information
on the chiral structure of new physics in $q\bar{q} \rightarrow t\bar{t}$.
For definiteness, we consider the two new observables:
\begin{eqnletter}
D &\equiv & \frac{\sigma(t_R\bar{t}_L) - \sigma(t_L\bar{t}_R)}
{\sigma(t_R\bar{t}_R) + \sigma(t_L\bar{t}_L) +
\sigma(t_L\bar{t}_R) + \sigma(t_R\bar{t}_L)}\,,
 \\[0.1cm]
D_{\rm FB} &\equiv &
D (\cos\hat\theta \geq 0) -  D (\cos\hat\theta \leq 0)
\end{eqnletter}
which involve the sum and difference of the coefficients
$C_1^\prime$ and $C_2^\prime$, respectively.
We found that $|D|$ and $|D_{\rm FB}|$ could be as large as $0.1$
in the region $|C_{1,2}^\prime\,(1\,{\rm TeV}/\Lambda)^2|\lesssim 1$
\cite{ko2}
Note that there are no experimental constraints on the
$D$ and $D_{\rm FB}$ observables yet.

%\bigskip 

\section{Possible underlying physics ?} 

Now we study specific new physics that could generate the 
relevant dim-6 operators with corresponding Wilson coefficients
(please refer to Refs.~\cite{ko1,ko2} the definitions of the interaction
lagrangians). 
It is impossible to exhaust all the possibilities, and we consider
the following interactions of quarks with spin-1 
flavor-conserving (FC) color-octet $V^a_{8A}$ vectors,
spin-1 flavor-violating (FV) color-singlet $\tilde{V}_{1A}$ and color-octet 
$\tilde{V}^a_{8A}$ vectors, and spin-0 FV color-singlet $\tilde{S_1}$
and color-octet $\tilde{S}^a_{8A}$ scalars ($A=L,R$).
After integrating out the heavy vector and scalar fields, we obtain 
the Wilson coefficients $C_{8q}^{AB}$ with $A,B=L,R$, explicit expressions
of which could be found in Refs.~\cite{ko1,ko2}. 

\begin{table}[t]
\caption{\label{tab:newparticles}
{\it
New particle exchanges and the signs of induced couplings
$C^{AB}$ ($A,B=R,L$),
$C_1-C_2$,
$C_1^\prime+C_2^\prime$, and
$C_1^\prime-C_2^\prime$.}
}
\begin{center}
\begin{tabular}{ccccccccc}
\hline %\hline
& & & & & & & &   \\[-0.3cm]
Resonance & $C^{RR}$ & $C^{LL}$ & $C^{LR}$ & $C^{RL}$ &
$C_1 - C_2$ &
$C_1^{'} + C_2^{'}$ & $C_1^{'} - C_2^{'}$ & $A_{\rm FB}$
\\[0.1cm]
\hline%\hline
& & & & & & & & \\[-0.3cm]
$\tilde{V}_{1R}$  &  $-$ & 0 & 0 & 0 &
$-$ & $-$ & $-$ & $\times$  \\[0.1cm]
$\tilde{V}_{1L}$  &  0 & $-$ & 0 & 0 &
$-$ & $+$ & $+$ & $\times$  \\[0.1cm]
$\tilde{V}_{8R}$  & $+$ & 0 & 0 & 0  &
$+$ & $+$ & $+$ & $\surd$ \\[0.1cm]
$\tilde{V}_{8L}$  & 0 & $+$ & 0 & 0  &
$+$ & $-$ & $-$ & $\surd$ \\[0.1cm]
\hline
& & & & & & & & \\[-0.3cm]
$\tilde{S}_{1R}$  & $0$ & $0$ & $0$ & $-$ &
$+$ & $+$ & $-$ & $\surd$ \\[0.1cm]
$\tilde{S}_{1L}$  & $0$ & $0$ & $-$ & $0$ &
$+$ & $-$ & $+$ & $\surd$ \\[0.1cm]
$\tilde{S}_{8R}$  & $0$ & $0$ & $0$ & $+$ &
$-$ & $-$ & $+$ & $\times$ \\[0.1cm]
$\tilde{S}_{8L}$  & $0$ & $0$ & $+$ & $0$ &
$-$ & $+$ & $-$ & $\times$ \\[0.1cm]
\hline
& & & & & & & & \\[-0.3cm]
$S_2^\alpha$ & $-$ & $0$ & $0$ & $0$ &
$-$ & $-$ & $-$ & $\times$ \\[0.1cm]
$S_{13}^{\alpha\beta}$  & $+$ & $0$ & $0$ & $0$ &
$+$ & $+$ & $+$ & $\surd$ \\[0.1cm]
\hline
& & & & & & & & \\[-0.3cm]
$V_{8R}$ & $\pm$ & $0$ & $0$ & $0$ &
$\pm$ & $\pm$ & $\pm$ & $\surd (+)$ or $\times (-)$ \\[0.1cm]
$V_{8L}$ & $0$ & $\pm$ & $0$ & $0$ &
$\pm$ & $\mp$ & $\mp$ & $\surd (+)$ or $\times (-)$ \\[0.1cm]
$V_{8R}\,,V_{8L}$ & indef. & indef. & indef. & indef. &
indef. & indef. & indef. & indef. \\[0.1cm]
\hline%\hline
\end{tabular}
\end{center}
\end{table}

\begin{figure}
\hspace{-0.5cm}
\includegraphics[width=8cm]{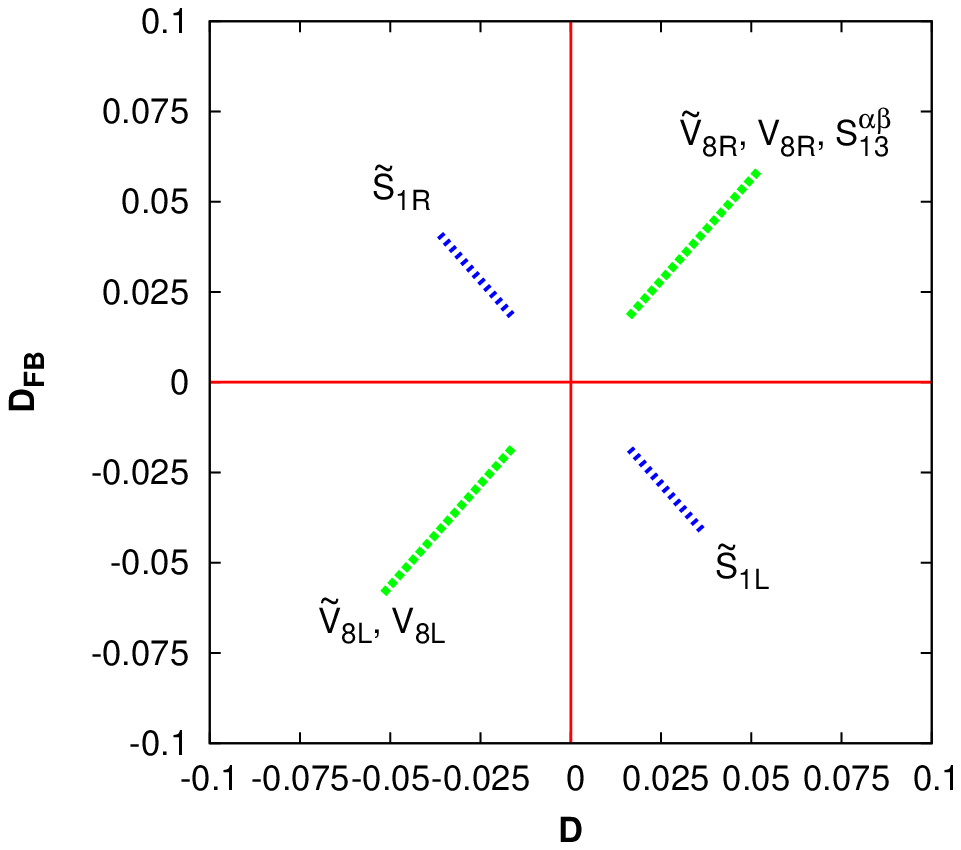}
\hspace{-1.5cm}
\includegraphics[width=8cm]{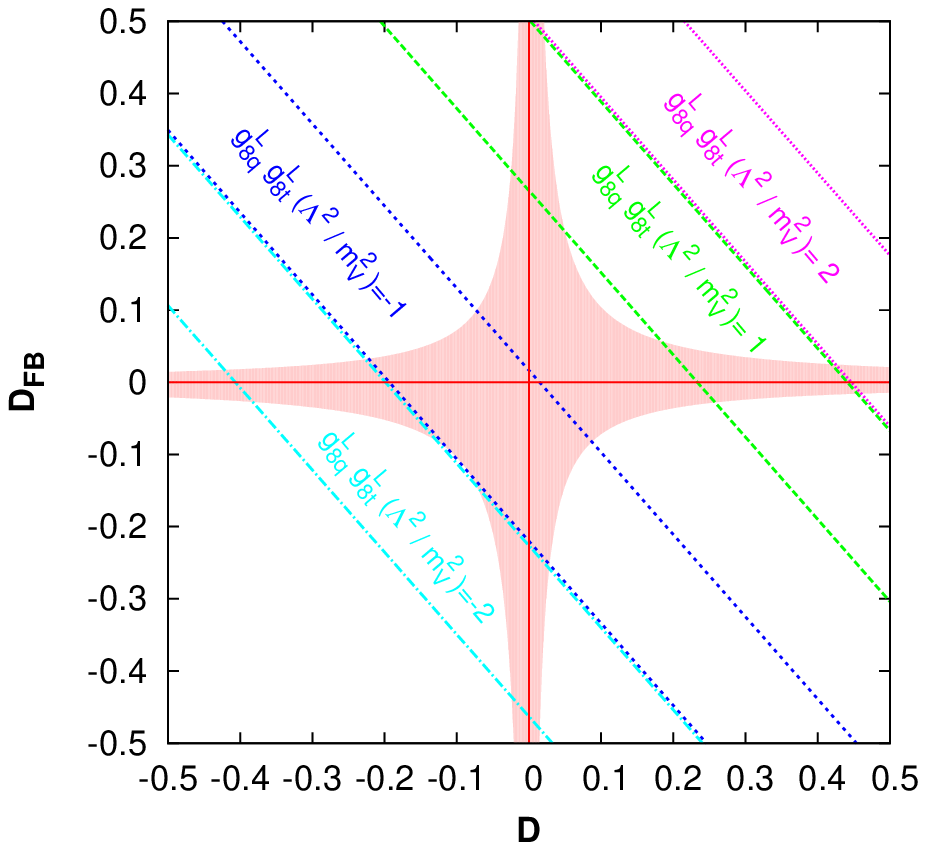}
\caption{\label{}
(a) The predictions for $D$ and $D_{\rm FB}$ of the models 
under consideration, being
consistent with the $\sigma_{t\bar{t}}$ and $A_{\rm FB}$ 
measurements at the 1-$\sigma$ level. We assume only one resonance
exists or dominates.
(b) The predictions for $D$ and $D_{\rm FB}$, being
consistent with the $\sigma_{t\bar{t}}$ and $A_{\rm FB}$
measurements at the 1-$\sigma$ level, for several values
of $g^L_{8q}g^L_{8t}\,(1\,{\rm TeV}/m_{V8})^2=$
$+2$ (magenta), $+1$ (green), $-1$ (blue), and $-1$ (sky blue), 
from the upper-right corner to the lower-left one.
The general model with
flavor-conserving color-octet
$V_{8R}$ and $V_{8L}$ vectors is considered. }
\label{fig:ddfb_onecoupling}
\end{figure}

Let us first consider the FV cases.
Among the FV interactions with vector or scalar bosons, 
$\tilde{V}_{8R,8L}$, $\tilde{S}_{1R,1L}$, and $S_{13}^{\alpha\beta}$ can give 
the correct sign for $(C_1 - C_2) \propto A_{\rm FB}$~\cite{ko1}.
But one can not discriminate one model from another only 
with the $A_{\rm FB}$ measurement. 
From Table~\ref{tab:newparticles}, we observe that each of the four cases
with $\tilde{V}_{8R}$, $\tilde{V}_{8L}$, $\tilde{S}_{1R}$, and $\tilde{S}_{1L}$ 
gives a different sign combination of $C_1^\prime+C_2^\prime$ and 
$C_1^\prime-C_2^\prime$.
In Fig.~\ref{fig:ddfb_onecoupling}, we show the prediction of each model for
$D$ and $D_{\rm FB}$ varying the model parameters 
which are consistent with the current measurements
of $\sigma_{t\bar{t}}$ and $A_{\rm FB}$ at the 1-$\sigma$ level. 
We observe that $D$ and $D_{\rm FB}$ take the same $(+,+)$
and $(-,-)$ signs for $\tilde{V}_{8R}$ and
$\tilde{V}_{8L}$, respectively, while they take
the different $(+,-)$ and $(-,+)$ signs for
$\tilde{S}_{1L}$ and $\tilde{S}_{1R}$, respectively.
The color-sextet scalar $S_{13}^{\alpha\beta}$ gives
the same $(+,+)$ sign as the $\tilde{V}_{8R}$ case.
Therefore, a simple sign measurement of $D$ and $D_{\rm FB}$ can 
endow us with the model-discriminating power.

Unlike the FV cases, the FC color-octet 
vectors can always accommodate the positive sign of $(C_1 - C_2)$.
For the case of $V_{8R}$ ($V_{8L}$), the couplings $g^R_{8q}$ ($g^L_{8q}$) 
and $g^R_{8t}$ ($g^L_{8t}$) must have different signs to 
accommodate the positive $A_{\rm FB}$. In Fig.~\ref{fig:ddfb_onecoupling}, 
we also show the predictions of the model with $V_{8R}$ or $V_{8L}$ vector
for $D$ and $D_{\rm FB}$.

%---------------------------------------------------------------------------------
\section{Beyond the effective field theory : the case of light $Z^{'}$}

Before closing this talk, let me make a few comments on the new physics 
scenarios with spin-1 objects, such as axigluon, $Z^{'}$, $W^{'}$ or 
$SU(3)_{u_R}$ flavor gauge bosons.  Whenever these new spin-1 particles have 
chiral couplings to the SM quarks, it is important to extend 
the SM Higgs sector too, in order that we can write (renormalizable)  
Yukawa couplings for all the SM fermions. One has to introduce new Higgs doublets 
that are charged under new gauge group, and they can affect the top FBA and the 
same sign top pair production rate in general. 
Also the new Higgs doublets can contribute to the $Wjj$ signals.  
These points were first noticed in Refs.~\cite{ko4,ko5}, and was presented 
by Chaehyun Yu in the poster session \cite{yu2} at this workshop. 
It is important to make sure that all the minimal ingredients  for the minimal 
consistent model are included before one starts phenomenological analysis. 

\section{Conclusions}

In this talk,  we considered the $t\bar{t}$ productions at the Tevatron 
using dimension-6 contact interactions relevant to 
$q\bar{q}\rightarrow t \bar{t}$, mainly concentrating on the 
top FBA, (FB) spin-spin correlation, and the $P$-odd 
longitudinal (anti)top polarization of $P_L$ and $\bar{P}_L$. 
The $P$-odd top-quark longitudinal polarization observables 
Both $P_L$ and $\bar{P}_L$ can be nonzero in many new physics 
scenarios for the top FBA, in sharp contrast to the case of pure QCD,
and can give another important clue for the chiral structure of new physics, 
which is completely independent of $\sigma_{t\bar{t}}$ or $A_{\rm FB}$. 
Our results in Table I and Fig.~2  encode the predictions for the $P$-odd 
observables corresponding to the polarization difference 
$(P_L - \bar{P}_L)$ in various new physics scenarios 
in a compact and an effective way, when those new particles
are too heavy to be produced at the Tevatron but still affect $A_{\rm FB}$.  
If these new particles could be produced directly at the Tevatron or 
at the LHC,  we cannot use the effective lagrangian any more. 
We have to study specific models case by case including the new particles
explicitly, and anticipate rich phenomenology at colliders as well as 
at low energy \cite{progress}. 

%\appendix

%\section{}
%Let us go then, you and I\ldots

\acknowledgments
I am grateful to Dong Won Jung, Jae Sik Lee, Yuji Omura and Chaehyun Yu
for enjoyable collaborations on the subjects presented in this talk.

\end{document}